\input epsf
\input psfig.sty
\documentclass[referee]{aa}
\thesaurus{07.16.1}
\title{Planetary migration in evolving planetesimals discs}
\author{S. Ye\c{s}ilyurt\inst{1}, A. Del Popolo\inst{1,2,3}, N. Ercan\inst{1} 
}
\offprints{A. Del Popolo - {\it E-mail adelpop@unibg.it}}
\institute{$^1$ Dipartimento di Matematica, Universit\`{a} Statale
di Bergamo,
via dei Caniana, 2 - I 24129 Bergamo, ITALY \\
$^2$ Feza G\"ursey Institute, P.O. Box 6 \c Cengelk\"oy, Istanbul, Turkey \\
$^3$ Bo$\breve{g}azi$\c{c}i University, Physics Department,
80815 Bebek, Istanbul, Turkey\\
}
\titlerunning{Planetary migration in evolving planetesimals discs}
\authorrunning{S. Ye\c{s}ilyurt et al. }
\date{}

\begin{document}
\maketitle
\begin{abstract}
In the current paper, we further improved the model for the migration of
planets introduced in Del Popolo et al. (2001) and extended to time-dependent
planetesimal accretion disks in Del Popolo and  Ek\c{s}i (2002). In the current
study, the assumption of Del Popolo and  Ek\c{s}i (2002), that the surface density
in planetesimals is proportional to that of gas, is released. 
In order to obtain the evolution of planetesimal density, 
we use a method developed in Stepinski and Valageas (1997) which 
is able to simultaneously follow the evolution of
gas and solid particles for up to $10^7 \mathrm{yrs}$. Then, the disk model
is coupled to migration model introduced in Del Popolo et al. (2001) in order
to obtain the migration rate of the planet in the planetesimal.  
We find that the properties of solids known to exist in protoplanetary
systems, together with reasonable density profiles for the disk, lead to a
characteristic radius in the range $0.03-0.2$ AU for the final semi-major
axis of the giant planet.
%
%
\end{abstract}

\section{Introduction}

The discovery of solar-like stars showing evidences for planets
orbiting around them (\cite{Mayor1}, \cite{Marcy1},
\cite{Vogt1}, \cite{Butler1}) has greatly intensified the
interest in understanding the formation and evolution of planetary
systems, as well as the long-standing problem of the solar system
origin. 

The extra-solar planets
discovered so far are all more massive than Saturn, and most either orbit
very close to their stars or travel on much more eccentric paths than any of
the major planets in our Solar System. 

It is difficult to explain the properties of these planets 
using the standard model for planet formation 
(\cite{Lissauer1}; \cite{Boss1}). Current theories (\cite{Mizuno1};
\cite{Bod1}) predict that giant planets were formed by gas accretion onto
massive ($ \simeq 15 M_{\rm \oplus}$) rocky cores which themselves were the
result of the accumulation of a large number of icy planetesimals. The most
favourable conditions for this process are found beyond the so-called
``snow line'' (\cite{Hayashi1}; \cite{Sasselov1}). As a consequence,
this standard model predicts nearly circular planetary orbits and giant
planets distances $\geq 1$ AU from the central star where the temperature
in the protostellar nebula is low enough for icy materials to condense
(\cite{Boss1}; \cite{Boss2}; but see also \cite{Wuchterl1};
\cite{Wuchterl2}).

%
%
Therefore, in the case of close-in giants, it is very unlikely that such
planets were formed at their present locations. Then, the most natural
explanation for this paradox, and for planets on very short orbits, is that
these planets have formed further away in the protoplanetary nebula and they
have migrated afterwards to the small orbital distances at which they are
observed (see DP1 and DP2 for a detailed discussion of migration mechanisms).
In particular, in DP1 and DP2 we showed that dynamical friction between a
planet and a planetesimals disk is an important mechanism for planet
migration and we pointed out that some advantages of the model are: \newline
a) Planet halt is naturally provided by the model. \newline
b) It can explain planets found at heliocentric distances of $>0.03-0.04$
AU, or planets having larger values of eccentricity. \newline
c) It can explain metallicity enhancements observed in stars having planets
in short-period orbits.\newline
d) Radial migration is possible with moderate masses of planetesimal disks,
in contrast with other models.

In DP1, following \cite{Opik1}, it was
assumed that the surface density in planetesimals $\Sigma _{\mathrm{s}}$
varies as $\Sigma _{\mathrm{s}}(r)=~\Sigma _{\odot }(1\mathrm{AU}/r)^{3/2}$,
where $\Sigma _{\odot }$, is the surface density at 1 AU.
In DP2 the previous assumption was substituted by a more reliable model for
the disk, and in particular we used a time-dependent accretion disk, since
it is widely accepted that the solar system at early phases in its evolution
is well described by this kind of structure. An important assumption of DP2
was that the surface density in planetesimals remains proportional to that
of gas: $\Sigma _{\mathrm{s}}(R,t)\propto \Sigma (R,t)$. However, it is
well-known that the distribution of planetesimals emerging from a turbulent
disk does not necessarily reflect that of gas (e.g., \cite{SV1}, %
\cite{SV2}). Indeed, in addition to gas-solid coupling, the evolution of
the distribution of solids is also godetermined by coagulation,
sedimentation and evaporation/condensation. In order to take into account
these effects we use the method developed in \cite{SV2} which is able to
simultaneously follow the evolution of gas and solid particles for up to $%
10^{7}\mathrm{yr}$. The main approximation used in this model is to
associate one grain size to a given radius and time. Then, we use the radial
distribution of planetesimals given by this model to evaluate the planet
migration, which is calculated as in \cite{DP1}.

This paper is organized as follows. In Sect. 2, 
we describe the disk model we use to obtain the 
distribution of the planetesimal. 
Then, in Sect. 3, we briefly review the
migration model introduced in \cite{DP1}. Finally, we describe our
results in Sect. 4 and Sect. 5 is devoted to conclusions.

\section{Disk model}



It is well-known that protostellar disks around young stellar objects are
common: between 25\% to 75\% of young stellar objects in the Orion nebula
seem to have disks with mass $10^{-3}M_{\odot }<M_{\mathrm{d}%
}<10^{-1}M_{\odot }$ and size $40\pm 20$ AU \cite{Beckwith1}. Moreover,
observations of circumstellar disks surrounding T Tauri stars support the
view of disks having a limited life-span and characterized by continuous
changes during their life.  
These evidences have led to a large consensus
about the nebular origin of the Solar System. Moreover, it clearly appears
necessary to model both the spatial and temporal changes of the disk (which
cannot be handled by the minimum-mass model nor by steady-state models).
Besides, one also needs to describe the global evolution of the solid
material which constitutes, together with the gas, the protoplanetary disk. 
\footnote{The knowledge of
this distribution and its time evolution is important to understand how
planets form and in this paper it is a key issue since we wish to study the
planet migration due to the interaction between planets and the local
distribution of solid matter.}
As usual, the time evolution of the surface density of the gas $\Sigma$ is
given by the familiar equation (e.g., \cite{SV2}): 
\begin{equation}
\frac{\partial \Sigma}{\partial t} -\frac{3}{r}\frac{\partial }{\partial r}%
\left[ r^{1/2}\frac{\partial }{\partial r}\left( r^{1/2}\nu _{t}\Sigma
\right) \right] =0  \label{gasevol}
\end{equation}
where $\nu_{\mathrm{t}}$ is the turbulent viscosity. Since $\nu_{\mathrm{t}}$
is not an explicit function of time, but instead depends only on the local
disk quantities, it can be expressed as $\nu_{\mathrm{t}}=\nu_{\mathrm{t}%
}(\Sigma,r)$ and Eq.(\ref{gasevol}) can be solved subject to boundary
conditions on the inner and outer edges of the disk. The opacity law needed
to compute $\nu_{\mathrm{t}}$ is obtained from \cite{Ruden1}. Then, 
Eq.(\ref{gasevol}) is solved by means of an implicit scheme. Note that the evolution
of the gas is computed independently from the evolution of particles (which
only make $\sim 1\%$ of the gas mass). Next, from $\Sigma(r,t)$ we can
algebraically find all other gas disk variables.

Next, as described in \cite{SV2} the evolution of the surface density of
solid particles $\Sigma_{\mathrm{s}}$ is given by: 
\begin{equation}
\frac{\partial \Sigma_{\mathrm{s}} }{\partial t} = \frac{3}{r} \frac{%
\partial }{\partial r} \left[ r^{1/2} \frac{\partial}{\partial r} (\nu_{%
\mathrm{s}} \Sigma_{\mathrm{s}} r^{1/2}) \right] + \frac{1}{r}\frac{\partial 
}{\partial r} \left[ \frac{2r \Sigma _{s} \langle \overline{v}_{\phi}
\rangle_{s}} {\Omega _{k}t_{s}} \right] .  \label{solidevol}
\end{equation}
The first diffusive term is similar to Eq.(\ref{gasevol}), where the
effective viscosity $\nu_{\mathrm{s}}$ is given by: 
\begin{equation}
\nu_{\mathrm{s}} = \frac{\nu}{\mathrm{Sc}} \hspace{0.4cm} \mbox{with} 
\hspace{0.4cm} \mathrm{Sc} = \left( 1+\Omega _{\mathrm{k}} t_{\mathrm{s}}
\right) \sqrt{1+\frac{\overline{\mathbf{v}}^2}{V_{\mathrm{t}}^2}} .
\label{Schmidt1}
\end{equation}
Here we introduced the Schmidt number Sc which measures the coupling of the
dust to the gas turbulence. We also used the relative velocity $\mathbf{v}$
between a particle and the gas, the turbulent velocity $V_{\mathrm{t}}$, the
Keplerian angular velocity $\Omega _{\mathrm{k}}$ and the so-called stopping
time $t_{\mathrm{s}}$. 
The average $\langle ..\rangle _{s}$ refers to
the vertical averaging over the disk height weighted by the solid density.
The dimensionless quantity $(\Omega _{\mathrm{k}}t_{%
\mathrm{s}})$ measures the coupling of the solid particles to the gas. 
Therefore, the evolution of the dust radial distribution can be
significantly different from the behaviour of the gas, depending on the
particle size (see \cite{SV1} for a detailed study).

The second advective term in Eq.(\ref{solidevol}) is due to the lack of
pressure support for the dust disk as compared with the gas disk. Thus, it
is proportional to the azimuthal velocity difference $\overline{v}_{\phi }$
between the dust and the gas. 
We refer the reader to \cite{SV2} for a more detailed presentation, see
also \cite{Kornet1}.

In this article we are mainly interested in the distribution of solids at
small radii hence we consider only one species of solid particles:
high-temperature silicates with $T_{\mathrm{evap}}=1350$ K and a bulk
density $\rho _{\mathrm{bulk}}=3.3$ g cm$^{-3}$. Thus, in our simplified
model we follow the evolution of three distinct fluids: the gas, the vapour
of silicates and the solid particles. In this way, we obtain the radial distribution of 
the planetesimal swarm
after $10^{7}$ yr. 
Of course, at these late times when planetesimals have
typically reached a size of a few km or larger, gravitational interactions
should play a dominant role with respect to coagulation. However, if these
interactions do not significantly affect the radial distribution of solids
(note that the radial velocity of such large particles due to the
interaction with the gas is negligible) we can still use the outcome of the
fluid approach described above to study the migration of giant planets, as
detailed below.

\section{Migration model}

In order to study the migration of giant planets, the model developed in DP1
is used (see also DP2). Since this model has already been described in these
two papers, we only recall here the main points. We consider a planet
revolving around a star of mass $M_{\ast }=1M_{\odot }$. The equation of
motion of the planet can be written as: 
\begin{equation}
\mathbf{\ddot{r}}=\mathbf{F}_{\odot }+\mathbf{R}
\end{equation}%
where the term $\mathbf{F}_{\odot }$ represents the force per unit mass from
the star, while $\mathbf{R}$ is the dissipative force (i.e. the dynamical
friction term--see \cite{Melita1}).

In order to take into account dynamical friction in a disk structure, we use  
Binney (1977) model:
\begin{equation}
\mathbf{R}=-k_{\parallel }v_{1\parallel }\mathbf{e_{\parallel }}-k_{\perp
}v_{1\perp }\mathbf{e_{\perp }}  \label{eq:dyn}
\end{equation}
where $\mathbf{e_{\parallel }}$ and $\mathbf{e_{\perp }}$ are two unit
vectors parallel and perpendicular to the disk plane and $k_{\parallel }$
and $k_{\perp }$ are given in Binney (1977) and DP1.

Since the damping of eccentricity and inclination is more rapid than radial
migration (\cite{Ida1}; \cite{Ida2}; DP1), we only deal with radial
migration and we assume that the planet has negligible inclination and
eccentricity ($i_{\mathrm{p}}\sim e_{\mathrm{p}}\sim 0$) and that the
initial distance to the star of the planet is $5.2$ $\mathrm{AU}$. 
We do not need to follow the evolution of the size
distribution of planetesimals
since for $m\ll M$ the frictional drag obtained 
does not depend on the mass $m$ of the planetesimals, because the
velocity dispersion only depends on the mass $M$ of the giant planet. 
We merely use the planetesimal density $\rho
_{\mathrm{s}}$ reached after $10^{7}$ yrs, assuming that the height of the
planetesimal disk does not evolve significantly.

\section{Results}

\label{Results}

\begin{figure}[tbp]
\centerline{\hbox{(a)
\psfig{figure=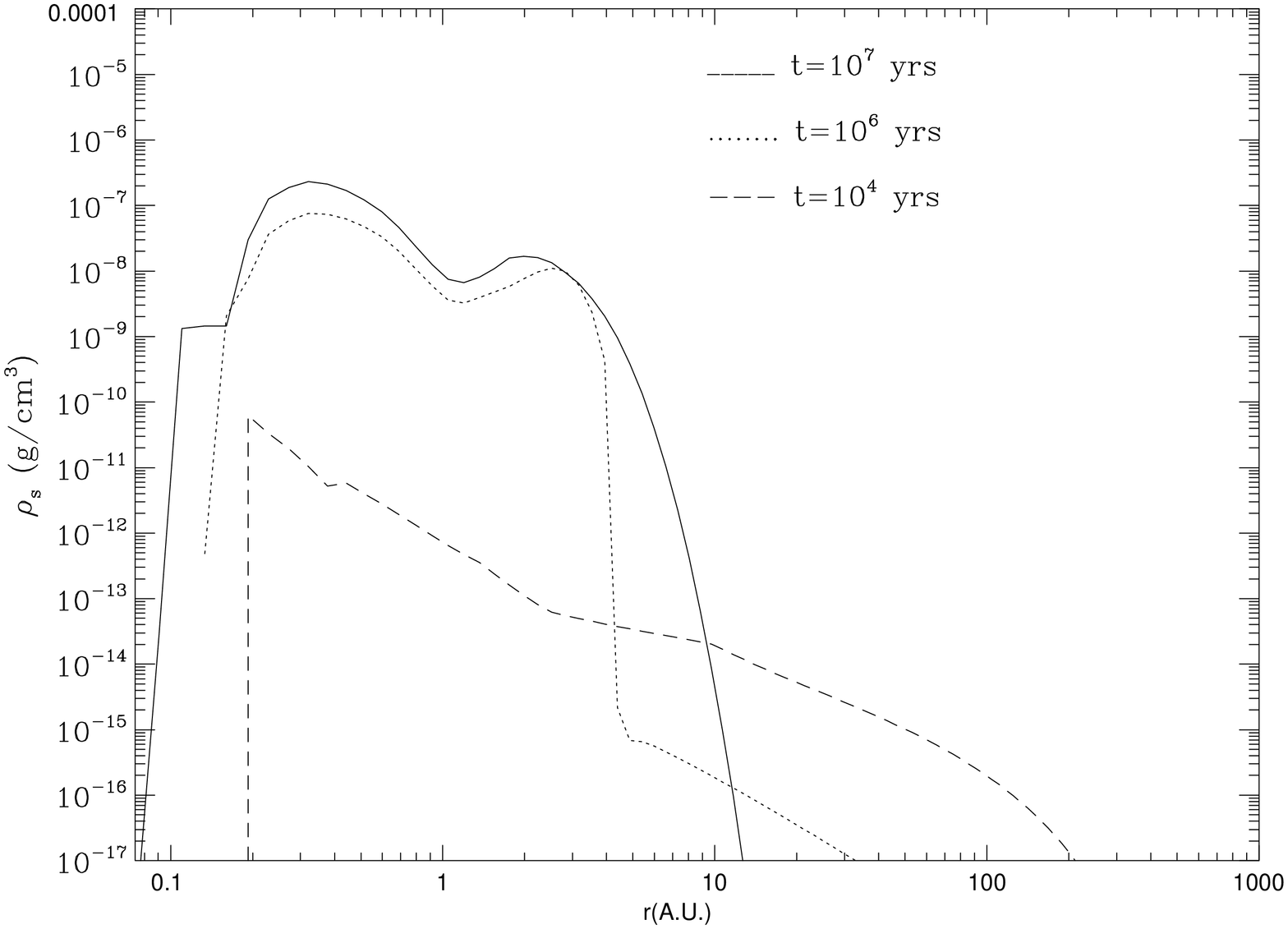,width=6cm,angle=270} (b)
\hspace{1cm}
\psfig{figure=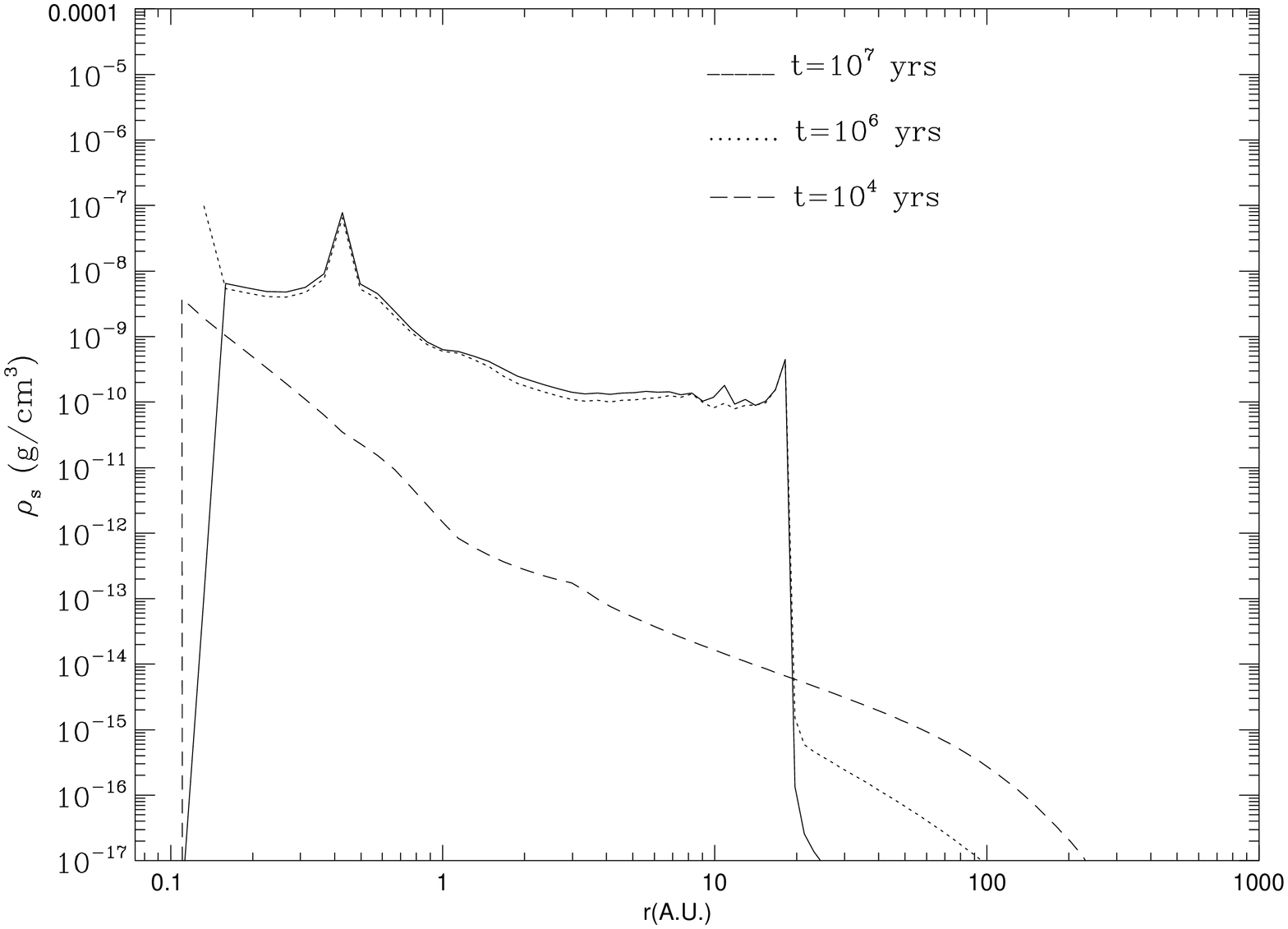,width=6cm,angle=270}
}}
\centerline{\hbox{(c)
\psfig{figure=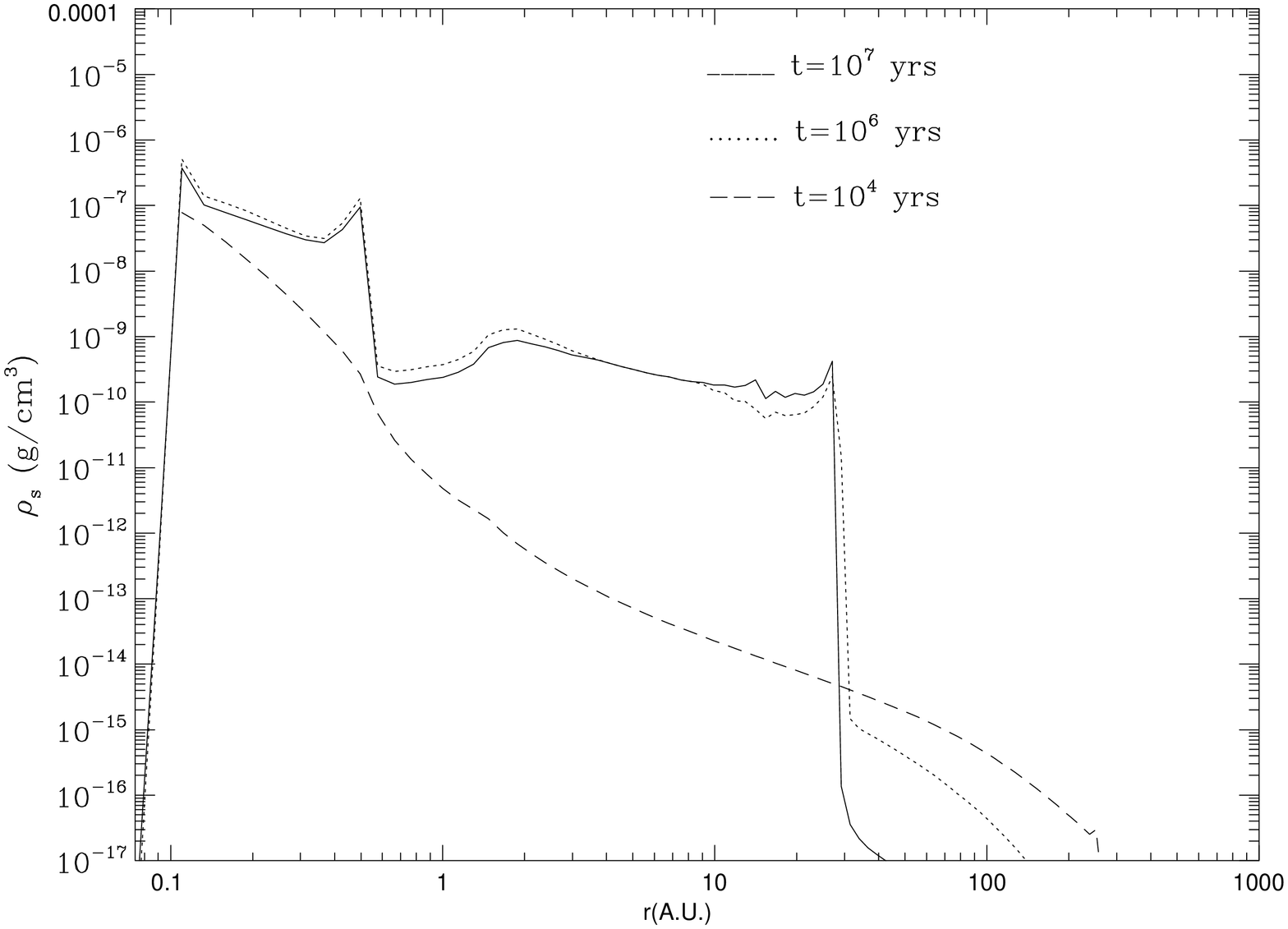,width=6cm,angle=270} (d)
\hspace{1cm}
\psfig{figure=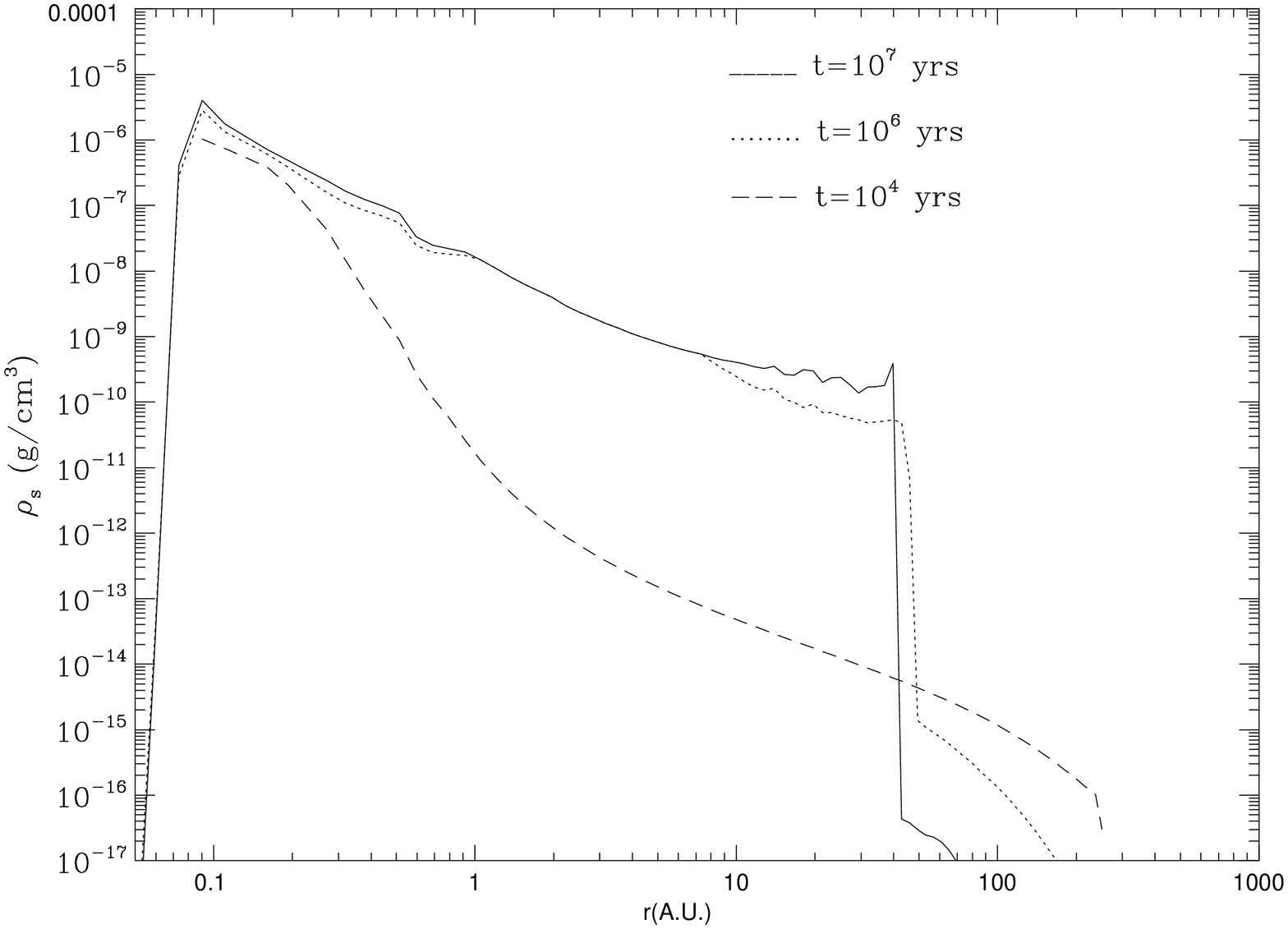,width=6cm,angle=270}
}}
\caption{(a) Evolution of solids for a disk with $M_d=0.1 M_{\odot}$ and $%
\protect\alpha=0.1$ at $t=10^4 \mathrm{yrs}$ (dashed line), $t=10^6 \mathrm{%
yrs}$ (dotted line) and $t=10^7 \mathrm{yrs}$ (solid line). (b) Same as
Fig. 1a but with $\protect\alpha=0.01$. (c) Same as Fig. 1a but with 
$\protect\alpha=0.001$. (d) Same as Fig. 1a but with $\protect\alpha=0.0001$}
\label{FigM1solid}
\end{figure}

In this article, similarly to DP1 and DP2, we are mainly interested in
studying the planet migration due to the interaction with planetesimals. Our
model starts with a fully formed gaseuos giant planet of 1 $M_{J}$ at 5.2 AU.
For this reason we assume that the gas is almost dissipated when the planet
starts its migration. 
\footnote{Clearly the effect of the presence of gas should be that of accelerating the
loss of angular momentum of the planet and to reduce the migration time.}
While
the gas tends to be dissipated, (several evidences show that the disk
lifetimes range from $10^{5}$ yr to $10^{7}$ yr, see \cite{Strom1}; %
\cite{Ruden1}), the coagulation process induces an increase of the
density of solid particles with time (see Fig. \ref{FigM1solid}) and gives rise to objects of increasing dimensions.

Once solids are in the form of planetesimals, the gas coupling becomes
unimportant and the radial distribution of solids does not change any more.
This is why we do not need to calculate its evolution for times longer than $%
10^{7}$ years. \footnote{Note however that the size distribution of planetesimals keeps changing due
to their mutual gravitational interaction.} 
Then the disk is populated by
residual planetesimals for a longer period. Here it is important to
emphasize that planetesimal formation is not independent from initial
conditions. In particular, the final solid surface density depends in an
intricate way on the initial disk mass $M_{d}$ and on the turbulent
viscosity parameter $\alpha $, (see \cite{SV2} and \cite{Kornet1}).

In order to investigate the dependence of the giant planet migration on the
properties of the protoplanetary disk we integrated the model introduced in
the previous sections for several values of the initial disk surface density
(i.e. several disk masses), and different values of $\alpha $. More
precisely, as in \cite{SV2} we consider an initial gas surface density of
the form: 
\begin{equation}
\Sigma _{0}(r)=\Sigma _{1}\left[ 1+(r/r_{1})^{2}\right] ^{-3.78}+\Sigma
_{2}(r/1\mathrm{AU})^{-1.5}.  \label{Sigma0}
\end{equation}%
The quantities $\Sigma _{1}$, $r_{1}$ and $\Sigma _{2}$ are free parameters
which we vary in order to study different disk masses. The values used are
given in Tab. 1, where $M_{d}$ is the gas disk mass (in units of $M_{\odot }$%
), $J_{50}$ is the disk angular momentum (in units of $10^{50}$ g cm$^{2}$s$%
^{-1}$), $\Sigma _{1}$ and $\Sigma _{2}$ are in g cm$^{-2}$ and $r_{1}$ is
in AU. The first term in Eq.(\ref{Sigma0}) ensures that there is some mass
up to large distances from the star, while the second term corresponds to
the central concentration of the mass and determine the location of the
evaporation radius. Note however that in any case the evaporation radius for
the high-temperature silicates we study here remains of order $0.1$ AU. As previoulsy 
explained, we consider only one species of solids:
high-temperature silicates with $T_{\mathrm{evap}}=1350$ K. We initialize
the dust subdisk at time $t=10^{4}$ yrs (i.e. after the gas distribution has
relaxed towards a quasi-stationary state) by setting the solid surface
density $\Sigma _{\mathrm{s}}$ as: $\Sigma _{\mathrm{s}}=6\times
10^{-3}\Sigma $ in order to account for cosmic abundance.

\begin{table}[tbp]
\caption{Properties of the initial gas disk}
\label{table1}
\begin{center}
\begin{tabular}{|l|l|l|l|l|}
\hline
$M_d$ & $J_{50}$ & $\Sigma_1$ & $r_1$ & $\Sigma_2$ \\ \hline
$10^{-1}$ & $911$ & $22$ & $200$ & $100$ \\ \hline
$10^{-2}$ & $85$ & $1.7$ & $200$ & $100$ \\ \hline
$10^{-3}$ & $5.5$ & $1.2$ & $50$ & $30$ \\ \hline
$10^{-4}$ & $0.46$ & $0.2$ & $50$ & $2.8$ \\ \hline
\end{tabular}%
\end{center}
\end{table}

We show in Fig. \ref{FigM1solid} the evolution
of the solid midplane density for the case $M_d=10^{-1} M_{\odot}$, 
and for different values of $\alpha$. We can see a converged
radial distribution of solids emerge at late times of order $10^6$ yr. Note
that although the radial distribution of planetesimals depends on the value
of $\alpha$, the total mass of solids in a disk is roughly independent of 
$\alpha$ since it remains approximately equal to the initial mass of solids
in the disk. This means that solids are reshuffled within the disk but they
are not lost into the star. \footnote{In fact, solids initially located close 
to the evaporation radius are lost
but they constitute a small percentage of the total solid material, which is
predominantly located in the outer disk.} The value of $\alpha$ determines
the radial distribution of solids: particles in a disk with a larger value
of $\alpha$ (more turbulent disk) have larger inward radial velocities and
consequently are locked into planetesimals closer to the star than particles
in a less turbulent disk. Thus, the smaller the value of $\alpha$ the
broader the final distribution of solids. The evaporation radius is located
at $\simeq 0.1 AU$ while the outer limit of converged $\rho_{\mathrm{s}}$ is
about $10$ AU (it moves outward for smaller values of $\alpha$, (e.g., $\simeq 70$ AU for 
$\alpha=10^{-4}$)). 

The coagulation process gives rise to solids of $10^6-10^7$ cm. In disks
characterized by smaller values of $\alpha$, and thus a more extended
distribution of solids, the range of sizes goes from $10^6-10^7$ cm at the
evaporation radius down to $10^3-10^4$ cm at the outer limit. This is because the coagulation process is less efficient at
larger radii where the solid density is smaller and the velocity dispersion
of the dust decreases (along with the gas temperature which governs the
turbulent velocity). 
At later times these solids will continue to increase
their sizes, but will not change their radial position, as they are already
large enough to have a negligible radial motion. 
We refer the reader to \cite{SV2} and \cite{Kornet1} for more detailed discussions of the
behaviour of protoplanetary disks.

\begin{figure}[tbp]
\centerline{\hbox{(a)
\psfig{figure=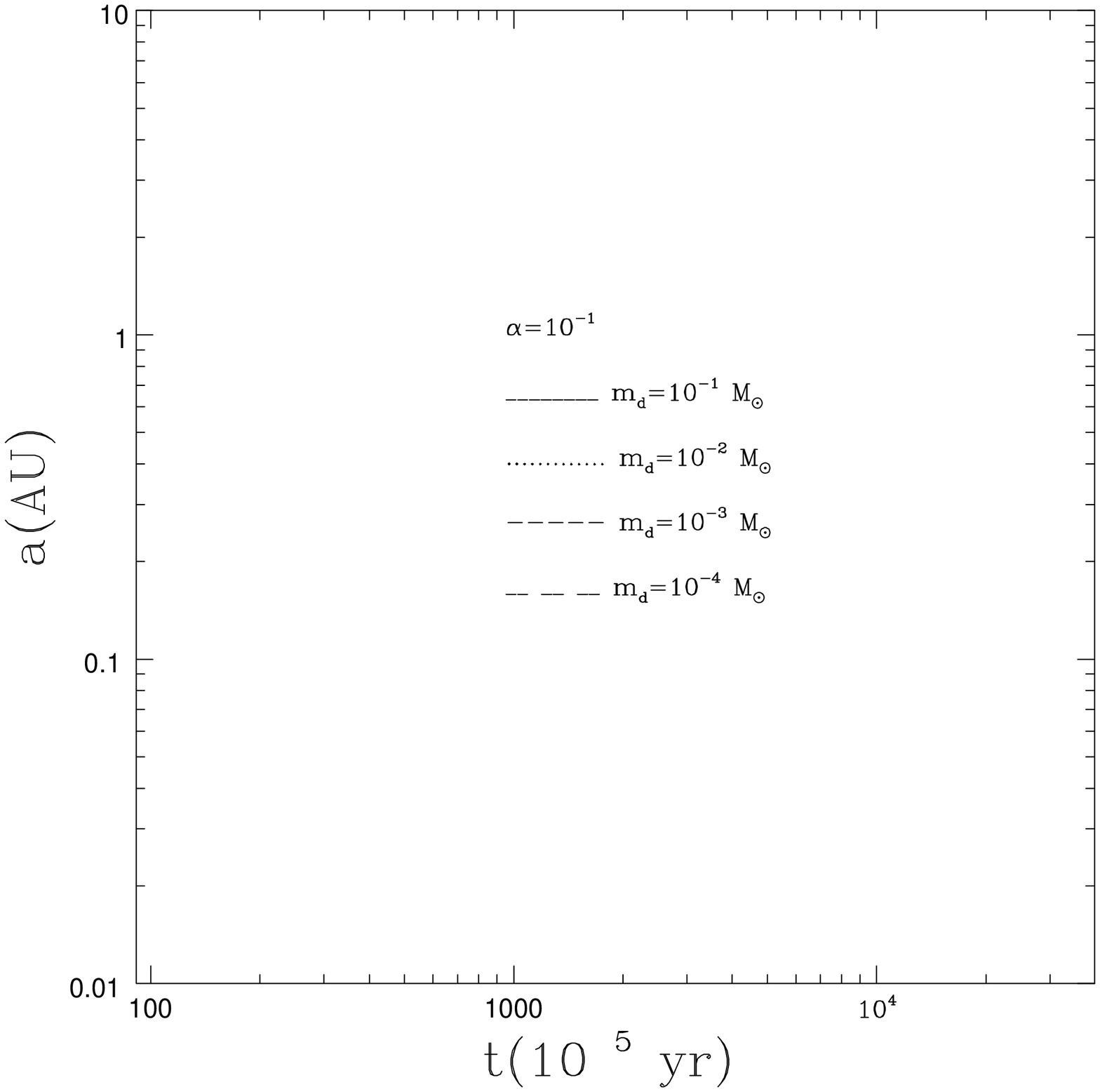,width=6cm} (b)
\vspace{1cm}
\psfig{figure=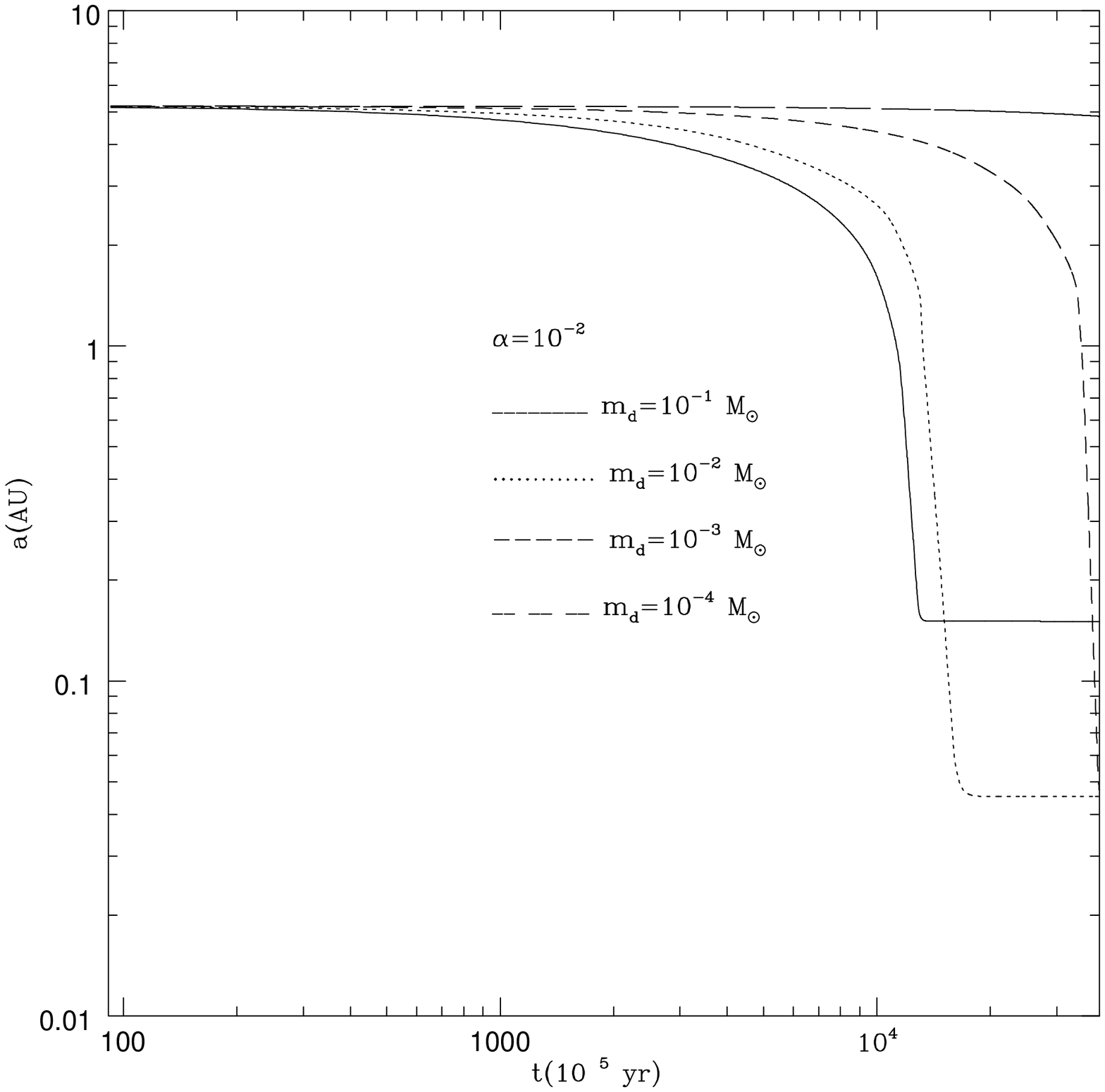,width=6cm}
}}
\centerline{\hbox{(c)
\psfig{figure=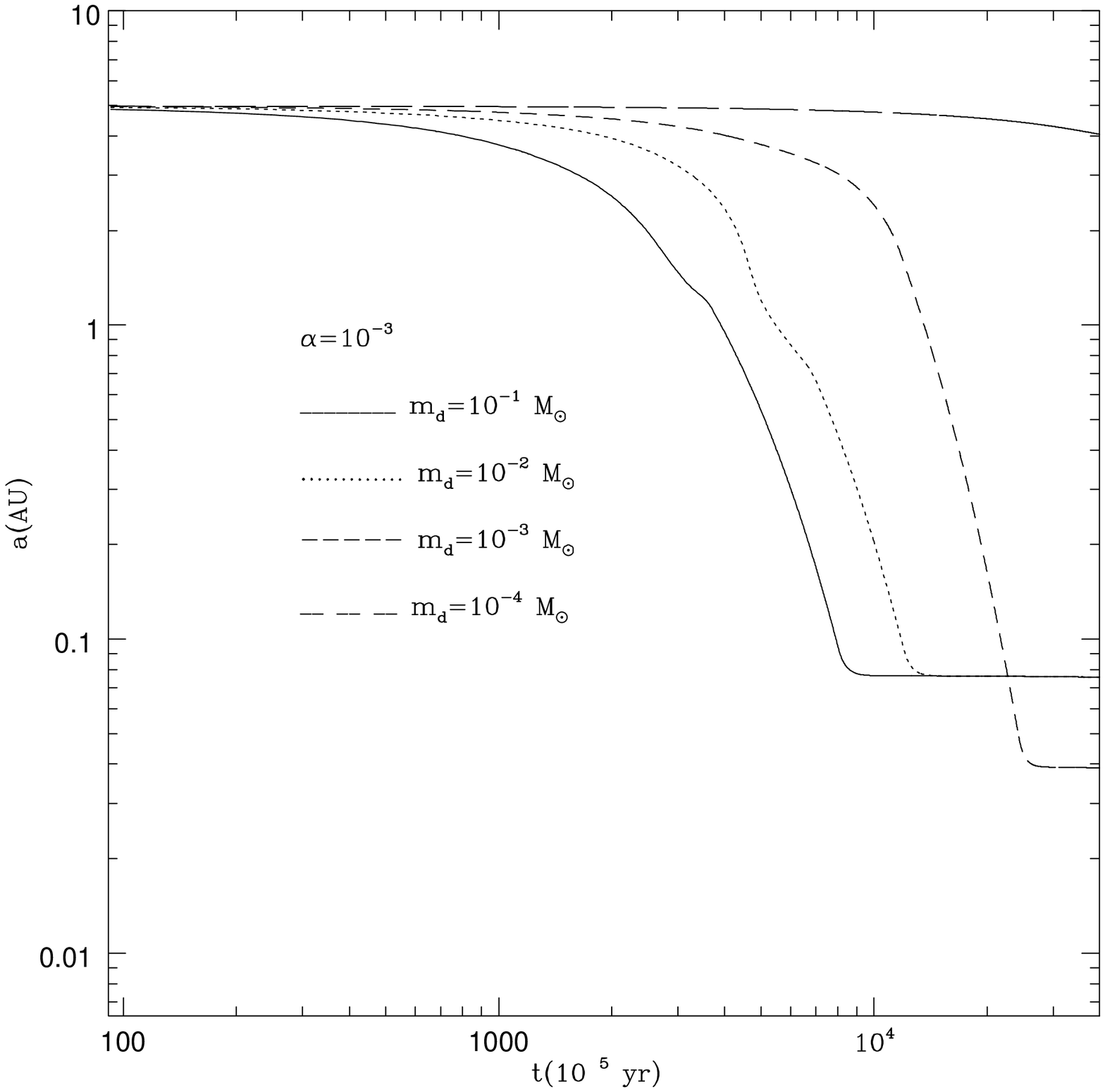,width=6cm} (d)
\vspace{1cm}
\psfig{figure=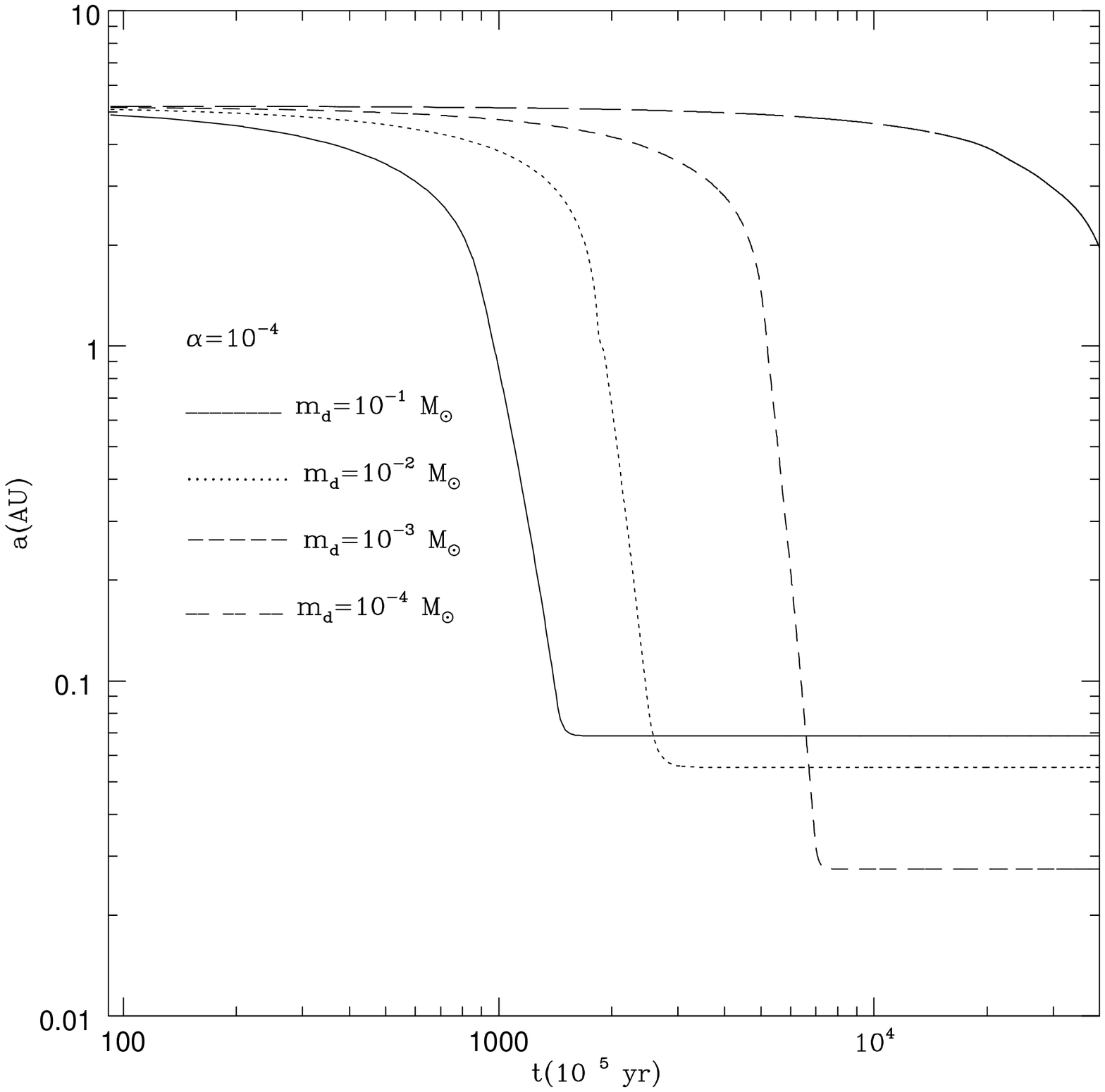,width=6cm}
}}
\caption{(a) The evolution of the semi-major axis $a(t)$ of a Jupiter-mass
planet, $M=1 M_{\mathrm{J}}$, in a planetesimal disk for $\protect\alpha=0.1$
and several values of $M_{\mathrm{d}}$, $0.1$ (solid line), $0.01$ (dotted
line), $0.001$ (short-dashed line) and $0.0001 M_{\odot}$ (long-dashed
line). (b) Same as Fig. (1a) but with $\protect\alpha=0.01$. (c) Same as
Fig. (1a) but with $\protect\alpha=0.001$. (d) Same as Fig. (1a) but with 
$\protect\alpha=0.0001$.}
\label{Figmig1}
\end{figure}

Now, using the converged radial distribution of planetesimals derived in the
previous section, we show in Fig. \ref{Figmig1} the evolution of the
semi-major axis $a(t)$ of a 1 $M_{J}$ planet in a disk for $\alpha =0.1$ and 
$M_{d}=0.1$ (solid line), $0.01$ (dotted line), $0.001$ (short-dashed line), 
$0.0001M_{\odot }$ (long-dashed line), respectively. We recall here that the
planet is initially located at $5.2$ AU with $i_{\mathrm{p}}\sim e_{\mathrm{p%
}}\sim 0$.

It can be clearly seen that for a fixed value of $\alpha $ a disk of larger
mass leads to a more rapid migration of the planet. This behaviour is quite
natural since a more massive disk obviously yields a stronger frictional
drag. Thus, in the cases $M=0.1,0.01M_{\odot }$ the planet migrates to $%
\simeq 0.08$ AU in $\simeq 1.5\times 10^{9}$ yr and to $\simeq 0.03$ AU in 
$\simeq 2.5\times 10^{9}$ yr, respectively. When the planet arrives at this
distance the dynamical friction switches off and its migration stops. The
stopping is simply due to the inner radius of the planetesimal disk. The
latter is set by the evaporation radius. Indeed, solid bodies cannot
condense at such small orbital radii $r\la 0.1$ AU because the temperature is
too high. Of course, this evaporation radius $r_{\mathrm{evap}}$ depends on
the properties of the solid grains we comsider. For instance, for ice
particules we have $r_{\mathrm{evap}}\sim 1$ AU (e.g., \cite{SV2}; 
\cite{Kornet1}). In this article we wish to understand the small orbital
radii of observed planets, over the range $0.03-0.15$ AU. Therefore, we are
interested in the inner regions of the disk where only high-temperature
silicates with $T_{\mathrm{evap}}\sim 1350$ K survive. This is why we
selected this component in this study. 
Then, the main point of Fig. \ref{Figmig1} is that the properties of solids known to exist in protoplanetary
systems, together with reasonable density profiles for the disk, lead to a
characteristic radius in the range $0.03-0.2$ AU for the final semi-major
axis of the giant planet. Note in particular that this process naturally
explains why the migration stops at such radii.

For less massive disks, $M_d=10^{-3}, 10^{-4} M_{\odot}$, the migration is
slower and the planet has not enough time to migrate belows $\simeq 4$ AU.
Note that we stop our calculation after $4 \times 10^9$ yr which is the
typical age of protoplanetary disks like ours (the Sun is $\sim 4.5 \times
10^9$ yr old). Moreover, the planetesimal disk should have cleared out by
this time. It is interesting to note that the planet moves closer to the
central star in the case $M=0.01 M_{\odot}$ than in the case $M=0.1
M_{\odot} $. This is due to the fact that less massive disks usually have a
smaller surface density which leads to a smaller temperature. This in turn
implies a smaller evaporation radius so that the planet can move closer to
the star. \footnote{Indeed, we have the energy balance: 
$T_{\mathrm{e}}^4 \propto \Sigma \nu_{\mathrm{t}} \Omega_k^2$ where $T_{\mathrm{e}}$ is the effective temperature, 
$\Sigma$ the gas surface density, $\nu_{\mathrm{t}}$ the turbulent viscosity
and $\Omega_k$ the Keplerian angular velocity. Besides, the mid-plane
temperature $T_c$ obeys the relation $T_c^4 \sim \tau T_{\mathrm{e}}^4$
where $\tau$ is the opacity. On the other hand, the opacity $\tau$ is given
by $\tau \sim \kappa \Sigma$ where $\kappa$ is the Rosseland opacity, while
the turbulent viscosity scales as 
$\nu_{\mathrm{t}} \sim \alpha T_c/\Omega_k$, hence we obtain: 
$T_c^3 \propto \kappa \alpha \Sigma^2$.} 
In order to
study the effect of viscosity on migration, we performed three other
calculations with $\alpha=10^{-2}, 10^{-3}$ and $10^{-4}$, also plotted in
Fig. \ref{Figmig1}. We can note that the dependence of the final radius on 
$\alpha$ is rather weak and non-monotonic because it competes with the
dependence on the surface density whose precise value is an intricate
function of $\alpha$. On the other hand, the migration time usually
increases going from $\alpha=10^{-4}$ up to $\alpha=0.1$. Indeed, as seen
from Fig. \ref{FigM1solid} a smaller $\alpha$ can
lead to a larger mid-plane solid density. This is partly due to the fact
that the dust disk height is smaller because the turbulence measured by 
$\alpha$ is weaker.

The results displayed in Fig. \ref{Figmig1} show that for $\alpha \la 0.01$ a
Jupiter-like planet can migrate to a very small distance from the parent
star, $0.03\mathrm{AU}<r<0.1\mathrm{AU}$, provided the disk mass is
sufficiently large $M_{d}\ga10^{-3}M_{\odot }$. Only in the cases 
$M_{d}\la 10^{-4}M_{\odot }$, or $M_{d}\la 10^{-3}M_{\odot }$ with $\alpha \ga 0.1$, the
interaction with the planetesimal disk is too weak to yield a significant
migration. 
Then, our results show that a final
radius of $0.03\mathrm{AU}<r<0.1\mathrm{AU}$ is a natural result, unless the
planetesimal disk has been cleared off too early on (e.g., by gravitational
scattering).

In summary, the present model predicts that, unless the disk mass is very
small $M_{d}\la10^{-4}M_{\odot }$, planets tend to move close to the parent
star and to pile up to distances of the order of $0.03-0.04$ AU (see Kuchner \& Lecar 2002). However,
with some degree of fine-tuning it is also possible to find a planet at
intermediate distances between its formation site and such small radii.

\section{Conclusions}

In the current paper, we further improved the model for the migration of
planets introduced in DP1 and extended to time-dependent planetesimal
accretion disks in DP2. After releasing the assumption of DP2 that the
surface density of planetesimals is proportional to that of gas, a simplified
model developed by Stepinski \& Valageas (1996, 1997), that is able to
simultaneously follow the evolution of gas and high-temperature silicates
for up to $10^{7}\mathrm{yr}$, is used. Then we coupled this disk model to
the migration model introduced in DP1 in order to obtain the migration rate
of the planets in the planetesimal disk and to study how the migration rate
depends on the disk mass, on its time evolution and on the dimensionless
viscosity parameter $\alpha $.

We found that in the case of disks having a total mass of 
$M_{d}>10^{-3}M_{\odot }$ planets can migrate inward over a large distance,
while if $M<10^{-3}M_{\odot }$ the planets remain almost at their initial
position. On the other hand, for $M_{d}\sim 10^{-3}M_{\odot }$ a significant
migration requires $\alpha \la10^{-2}$.

If the migration is efficient the planet usually ends up at a small radius
in the range $0.03-0.1$ AU which is simply set by the evaporation radius of
the gaseous disk which gave rise at earlier times to the radial distribution
of the planetesimal swarm. Therefore, our model provides a natural
explanation for the small observed radii of extra-solar giant planets. 
In order to
inhibit this process (so that Jupiter-like planets like our own remain at
larger distances $\ga5$ AU) the planetesimal disk must be cleared off over a
time-scale of the order of or smaller than $10^{9}$ yr (depending on the
properties of the disk) or the disk mass must be rather small (i.e. smaller
than $1M_{\mathrm{J}}$) which could suggest an alternative formation
scenario for such a giant planet (i.e. not related with the disk itself).

\end{document}